\pdfoutput=1

\def\dzero{D0}
\def\bfx{\mathbf{x}}
\def\bfy{\mathbf{y}}
\def\evt{{\mathrm{evt}}}
\def\sig{{\mathrm{sig}}}
\def\bkg{{\mathrm{bkg}}}
\def\perm{{\mathrm{perm}}}
\def\jets{{\mathrm{jets}}}
\def\JES{{\mathrm{JES}}}
\def\tbar{\bar t}
\def\mtbar{m_{\tbar}}
\def\msum{m_\mathrm{sum}}
\def\ttbar{t \bar t}
\def\progname#1{{\sc\lowercase{#1}}}
\def\mathunit#1{\mathrel{\mathrm{#1}}\mathclose{}\mathord{}}
\def\fb{\mathunit{fb}}
\def\ifb{\ensuremath{\fb^{-1}}}
\def\gev{\mathunit{Ge\kern -0.1em V}}
\def\pt{\ensuremath{{p_{\scriptscriptstyle T}}}}
\def\ra{\rightarrow}
\def\statsyst#1#2{\pm#1\ \textnormal{(stat.)}\pm#2\ \textnormal{(syst.)}}
\def\mete{\setbox0=\hbox{$E$}%
          \hbox{$E$\rlap{\kern -0.45em\raise0.08\ht0\hbox{/}}}}
\def\met{\ensuremath{\mete_T}}

\newdimen\figsize

\long\def\threeboxesgap#1#2#3#4{%
  \setlength\figsize{\hsize}%
  \addtolength\figsize{-#4}%
  \addtolength\figsize{-#4}%
  \divide\figsize by 3
  \vbox{%
  \makebox{\parbox[t]{\figsize}{\vskip 0.1pt #1}%
           \hspace{#4}%
           \parbox[t]{\figsize}{\vskip 0.1pt #2}%
           \hspace{#4}%
           \parbox[t]{\figsize}{\vskip 0.1pt #3}}}}
\long\def\threeboxes#1#2#3{\threeboxesgap{#1}{#2}{#3}{\columnsep}}

\documentclass[
    ,final            
  ]
  {aipproc}

\layoutstyle{6x9}

\begin{document}

\def\fmstate{{\small Work supported by the U.S. Department of Energy under contract No. DE-AC02-07CH11359.}}
\DeclarePagestyle{fermi}{}{\fmstate}{}{\fmstate}
\thispagestyle{fermi}

\title{Measurements of the Top Quark Mass at \dzero}


\author{Scott S. Snyder\\\small For the \dzero\ Collaboration}{
  address={Brookhaven National Laboratory, Upton, NY 11973, USA}
}

\begin{abstract}

We present measurements of the top quark mass based on
$3.6\ifb$ of data collected by the \dzero\ experiment during Run II of the
Fermilab Tevatron collider. We present results in the dilepton and
lepton+jets final states. We also present the measurement of the
mass difference between $t$ and $\bar t$ quarks observed in lepton+jets
final states of $\ttbar$ events in $1\ifb$ of data.

\end{abstract}

\maketitle


\section{Introduction}

The unique place in the Standard Model filled by the top quark
makes precise measurements of its mass of great interest.
We report preliminary top quark mass measurements
using $3.6\ifb$ of data
from the \dzero\ experiment~\cite{d0} at the Fermilab Tevatron,
as well a direct measurement of
the mass difference between the top quark and its antiparticle.

\section{Measurements of the top quark mass}

The most precise top quark mass measurements at \dzero\ are currently
obtained by the matrix element method~\cite{matrix-element}.
In this method, a likelihood is assigned to each event of the form
\begin{equation}
  P_\evt(\bfx; m_t, f_t) =
    \left[ f_t P_\sig(\bfx;m_t) + (1-f_t) P_{\,\bkg}(\bfx)\right],
\end{equation}
where $\bfx$ represents the full set of measured kinematic quantities
of the event, $m_t$ the top quark mass,
and $f_t$ is the $\ttbar$ signal fraction in the sample.
The $\ttbar$ signal likelihood is then
\begin{equation}
  P_\sig(\bfx;m_t) = {1\over\sigma'(m_t)} \int_{q_1,q_2,\bfy}
  \sum_{\scriptstyle \perm,\atop\scriptstyle\mathrm{flavors}} w_\perm\, dq_1\, dq_2 f(q_1) f(q_2)
  {(2\pi)^4|{\cal M}|^2 \over 4\sqrt{(q_1\cdot q_2)^2}}
  d\Phi_6 W(\bfx,\bfy),
\label{like}
\end{equation}
where $\bfy$ is the set of parton-level kinematic variables defining
the final state of the hard scatter, $q_i$ are the momenta of the
incoming partons, $f$ is the parton distribution function,
$\sigma'$ is the observable cross section
(including the detector efficiency), $\cal M$ is the matrix element
for the hard scatter, taken from an analytic calculation~\cite{mahlon-parke},
and $W$ is the transfer function from a
hard scatter to the measured variables.  The sum is over flavors
of incoming partons and assignments
of jets to final-state partons; if $b$-tagging is used,
a weight $w_\perm$ is applied
based on this assignment.  $P_{\,\bkg}$ is similar, except that
there is no dependence on $m_t$ and $\cal M$ is evaluated using
\progname{vecbos}~\cite{vecbos}.  An estimate
of $m_t$ and its uncertainty is then extracted from the joint likelihood
of a data sample via maximum likelihood.  Due to approximations present
in $\cal M$ and $W$, the mass extracted by this procedure will be systematically
biased.  The analysis is carried out on many simulated experiments with various
input $m_t$; the results are used to correct the extracted $m_t$ and
its uncertainty.  \progname{Alpgen}~\cite{alpgen} is used to
model the $\ttbar$ signal.

This measurement has been carried out in the $\ttbar\ra e\mu\nu\nu b\bar b$
channel with $3.6\ifb$
of data~\cite{emu36}.  Events are selected with an isolated, high-$\pt$
electron and muon
and two jets; there is also a requirement made on
$H_T = \pt(\ell_1) + \pt(j_1) + \pt(j_2)$.  This yields 154 candidate events.
The expected background fraction is about $16\%$ and is predominantly
due to $Z\ra\tau\tau$.  The fit to the
final $2.5\ifb$ subset of the data  is shown in Fig.~\ref{dilres}.

\begin{ltxfigure}
\parbox[t]{2.7in}{%
  \centering
  \includegraphics[width=5.5cm]{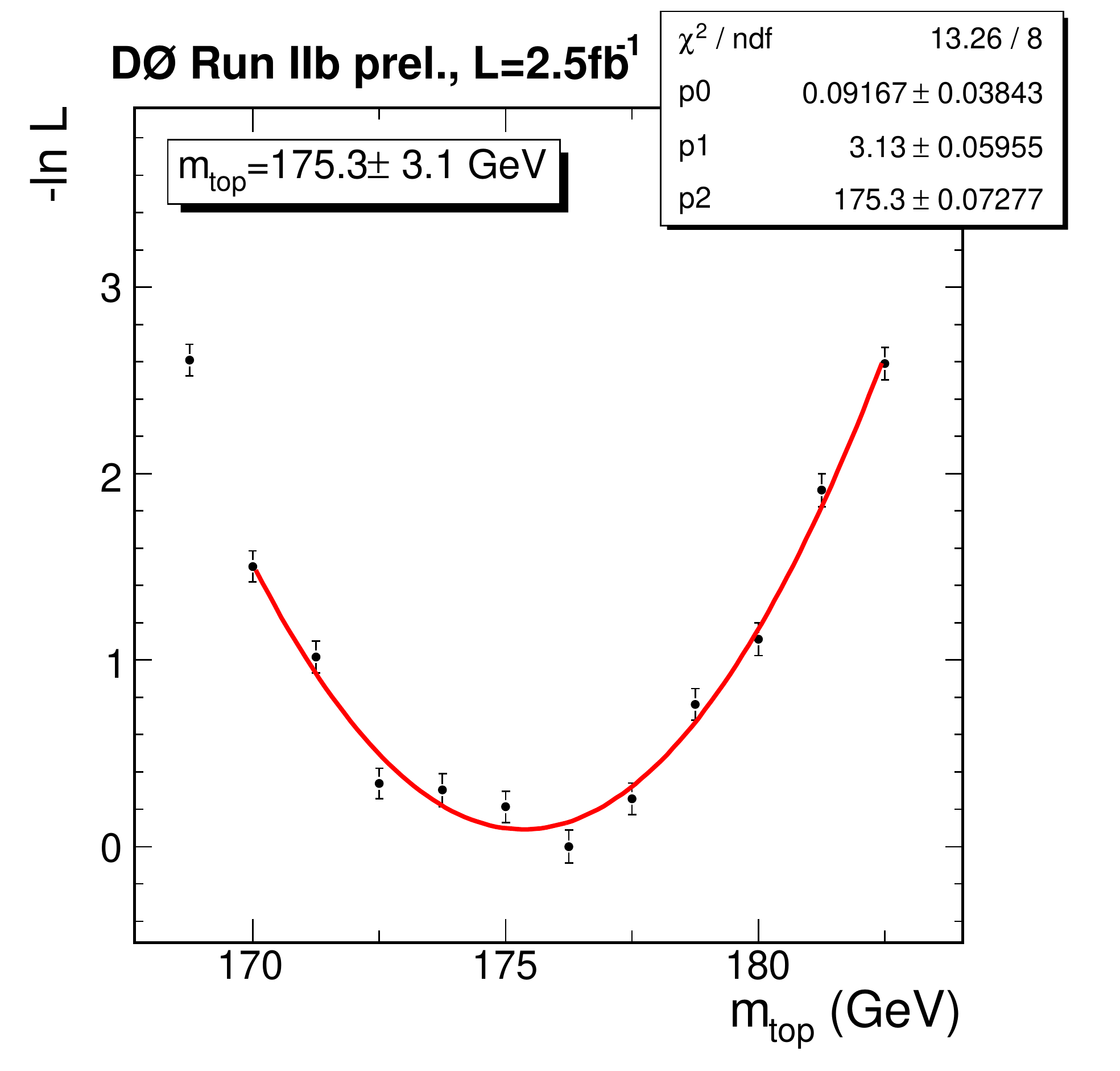}%
\caption{Fit for $m_t$ to $2.5\ifb$ of $e\mu$ data.  (Uncalibrated.)}
\label{dilres}
}%
\qquad
\parbox[t]{2.7in}{%
  \centering
  \includegraphics[width=6cm]{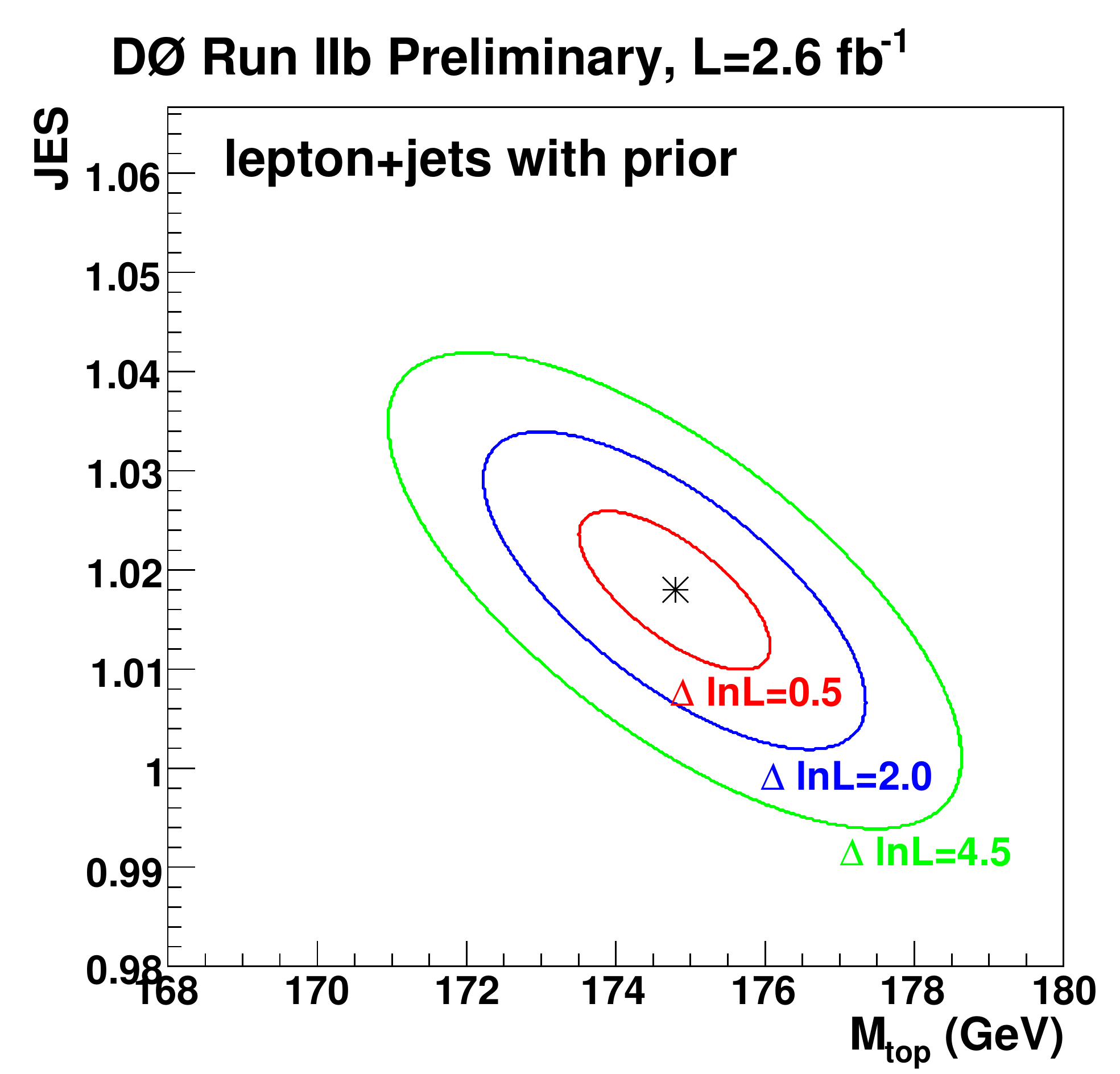}%
\caption{Fit for $m_t$ and JES to $2.6\ifb$ of $\ell+\jets$ data.
(Uncalibrated.)}
\label{ljetsres}
}%
\end{ltxfigure}

The result after calibration is $m_t = 174.8\statsyst{3.3}{2.6}\gev$.
Combined with previous \dzero\ results from channels in which both top quarks
decay to leptons~\cite{olddil},
the result is $m_t = 174.7\statsyst{2.9}{2.4}\gev$.

The analysis has also been performed in the $\ttbar\ra\ell\nu b\bar b qq$
(``$\ell+\jets$,'' where $\ell\equiv e, \mu$) channels
with $3.6\ifb$ of data~\cite{ljets36}.  Events are required to have
exactly one isolated, high-$\pt$ electron or muon, large $\met$,
and exactly four jets, at least one of which must be identified
as a $b$-jet.  This yields 615 events with a
background fraction of about $30\%$.  Each signal event has
a $W\ra jj$ decay;  by constraining this to the known mass of the $W$~boson,
the jet energy scale uncertainty may be reduced.  This is implemented
with an additional fit parameter JES, which is a multiplicative
scale factor on the jet energies.  A prior probability distribution
for JES is included, corresponding to the results of the standard
jet energy calibration.  Results for the final $2.6\ifb$ of data
are shown in Fig.~\ref{ljetsres}.

The result is $m_t = 174.7\statsyst{0.8}{1.6}\gev$.
Combining all top quark mass measurements from \dzero~\cite{d0combination}
yields $m_t = 173.2\statsyst{0.9}{1.5}$.
The precision of this measurement is now better than $1\%$;
also note that it is now systematics dominated.

\begin{ltxtable}
  \centering
  \parbox[t]{2.7in}{%
  \caption{Systematic uncertainties (in GeV) for the top
    quark mass measurements~\cite{tevcombination}.}
  \label{masssyst}
  \begin{tabular}{lcc}
        \hline
        \tablehead{1}{l}{b}{Source} &
        \tablehead{1}{c}{b}{$\ell\ell$} &
        \tablehead{1}{c}{b}{$\ell+\jets$} \\
        \hline
        \emph{In-situ} JES            & ---       & $\pm0.47$\\
        Jet $e/h$, $b$-tag, recon.
                                      & $\pm1.32$ & $\pm0.91$\\
        $b$ modeling                  & $\pm0.26$ & $\pm0.07$\\
        JES control samples           & $\pm1.46$ & $\pm0.84$\\
        Lepton momentum scale         & $\pm0.32$ & $\pm0.18$\\
        Signal model                  & $\pm0.65$ & $\pm0.45$\\
        Monte Carlo model diffs.      & $\pm1.00$ & $\pm0.58$\\
        Bkg.\ model (excl.\ QCD)      & $\pm0.08$ & $\pm0.08$\\
        Fit method + QCD bkg.         & $\pm0.51$ & $\pm0.21$\\
        Color reconnection            & $\pm0.40$ & $\pm0.40$\\
        Luminosity profile            & $\pm0.00$ & $\pm0.05$\\
        \hline
        Total:                        & $\pm2.43$ & $\pm1.60$\\
        \hline
  \end{tabular}
  }%
  \qquad
  \parbox[t]{2.7in}{%
  \caption{Systematic uncertainties (in GeV) for the top
    quark mass difference measurement.}
  \label{mdiffsyst}
  \centering
  \begin{tabular}{llc}
        \hline
        \tablehead{2}{l}{b}{Source} &
        \tablehead{1}{c}{b}{}\\
        \hline
        \multicolumn{2}{l}{{\it Physics modeling:}}&\\
        &Signal          & $\pm0.85$\\
        &PDF uncertainty & $\pm0.26$\\
        &Other           & $\pm0.14$\\
        \multicolumn{2}{l}{{\it Detector modeling:}}&\\
        &Jet resolution           & $\pm0.39$\\
        &Overall jet energy scale & $\pm0.08$\\
        &Wrong sign leptons       & $\pm0.07$\\
        &$b\bar b$ response asymmetry & $\pm0.42$\\
        &Other                        & $\pm0.22$\\
        \multicolumn{2}{l}{{\it Method:}}&$\pm 0.53$\\
        \hline
        \multicolumn{2}{l}{Total:}           & $\pm1.22$\\
        \hline
  \end{tabular}
  }%
\end{ltxtable}

Systematic uncertainties are summarized
in Table~\ref{masssyst}.  The categories shown in this table are the
result of a recent effort to make the evaluation of systematics
consistent across all Tevatron $m_t$ measurements.  Compared with previous
analyses, two new systematics have now been evaluated. 
``Color reconnection'' arises from variations
in the description of color reconnection of final-state
particles~\cite{color-reconn}.  ``Luminosity profile''
quantifies the uncertainty due to mismodeling the number
of collisions per bunch crossing.

When these results are combined with those from CDF,
the new world average is
$m_t = 173.1\statsyst{0.6}{1.1}\gev$~\cite{tevcombination},
for an overall precision of $0.75\%$.

\section{Top-antitop quark mass difference measurement}

CPT is generally believed to be a good symmetry nature; by the
CPT~theorem~\cite{CPT}, it holds for any local Lorentz-invariant
quantum field theory.  Nevertheless, it is important to search
for any violations.  One implication of CPT invariance is that the
mass of a particle and its antiparticle must be equal.  While many
precise such measurements have been made~\cite{cptmeas},
those for quarks have always been indirect.  The top quark is, however, unique:
due to its large mass, it decays before hadronization effects become
important.  Thus, the masses of the top and antitop quarks can be measured
directly and separately.

\dzero\ has now performed the first direct measurement of the mass difference
between a quark and its antiquark~\cite{mdiff}.
The measurement is performed in the
$\ell+\jets$ channel using the 2002--2006 data set, comprising about $1\ifb$.

The method is an extension of the matrix element method described earlier.
The matrix element of~\eqref{like} is modified so that the top and antitop
quark masses are specified separately.  For this analysis, the jet energy
scale factor JES is fixed to that determined from the mass measurement.
Thus, instead of $(m_t, \JES)$, we now extract $(m_t, \mtbar)$;
or equivalently $(\Delta m, \msum) = (m_t-\mtbar, (m_t+\mtbar)/2)$.
The leptonically-decaying top quark in each event is identified
as either a quark or antiquark based on the sign of the lepton.
A modified version of \progname{pythia}~\cite{pythia} is used to
simulate signal events with $m_t\ne \mtbar$.

\begin{figure}
  \centering
  \threeboxes{\includegraphics[width=\figsize]{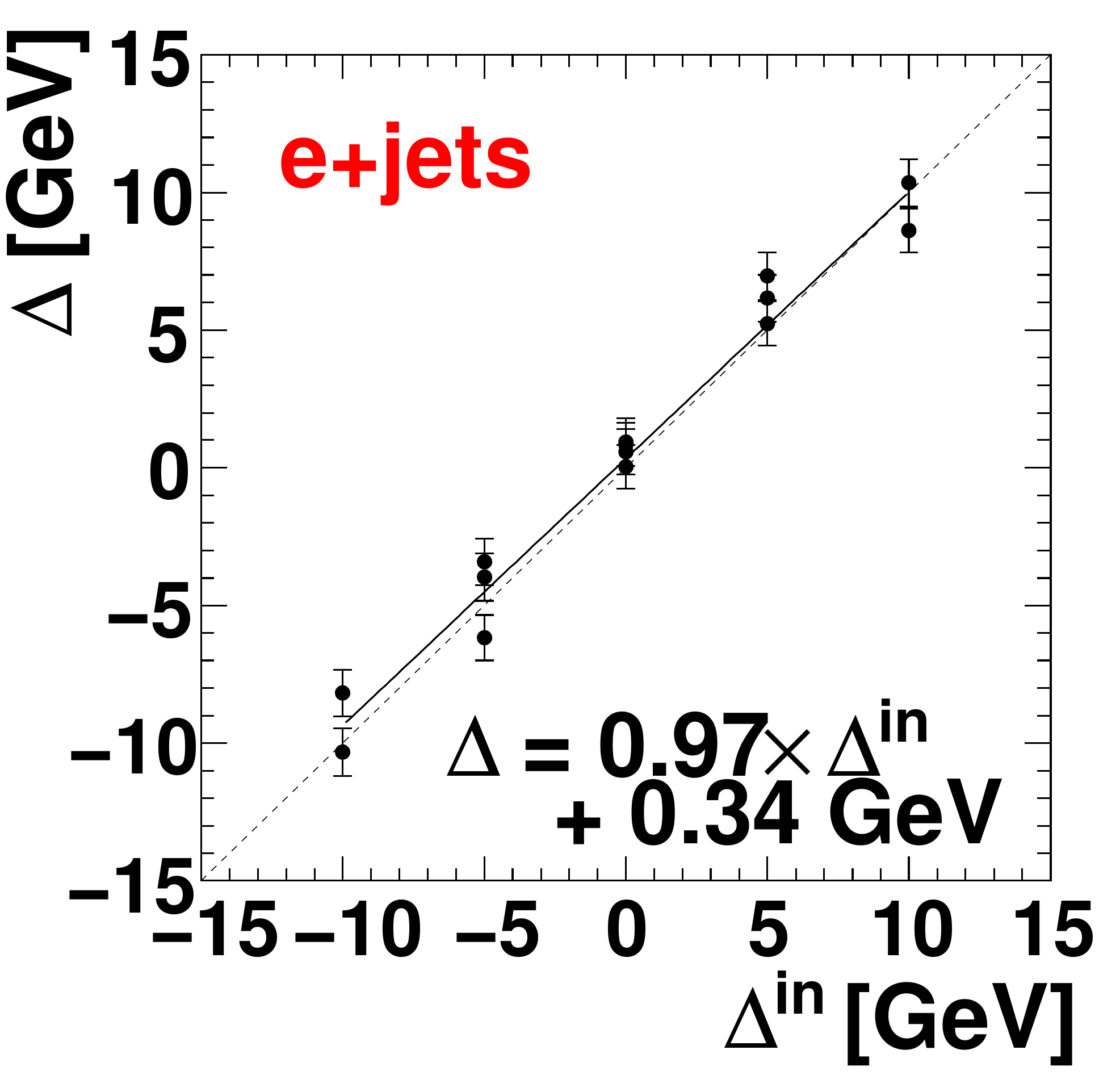}}%
             {\includegraphics[width=\figsize]{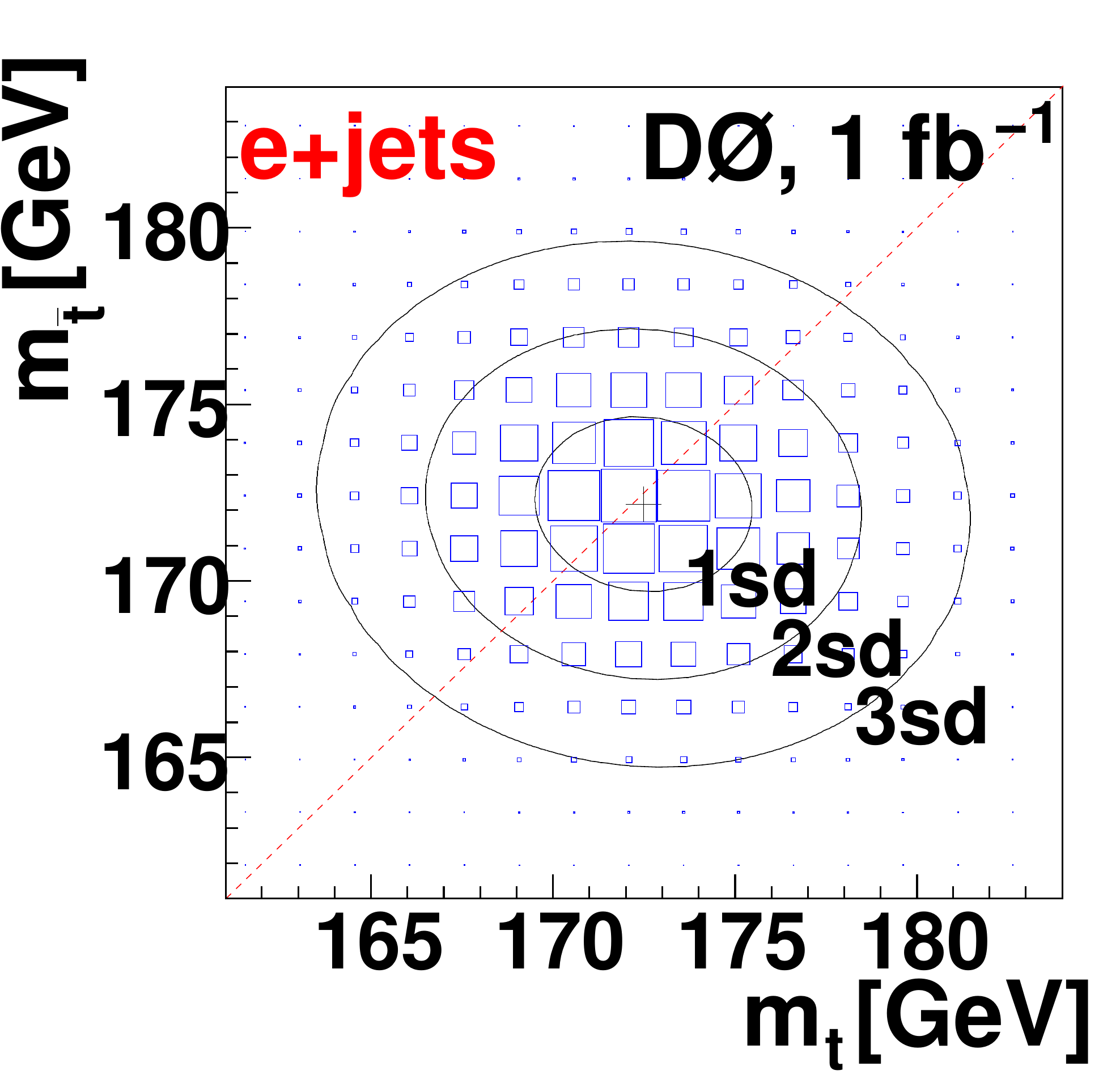}}%
             {\includegraphics[width=\figsize]{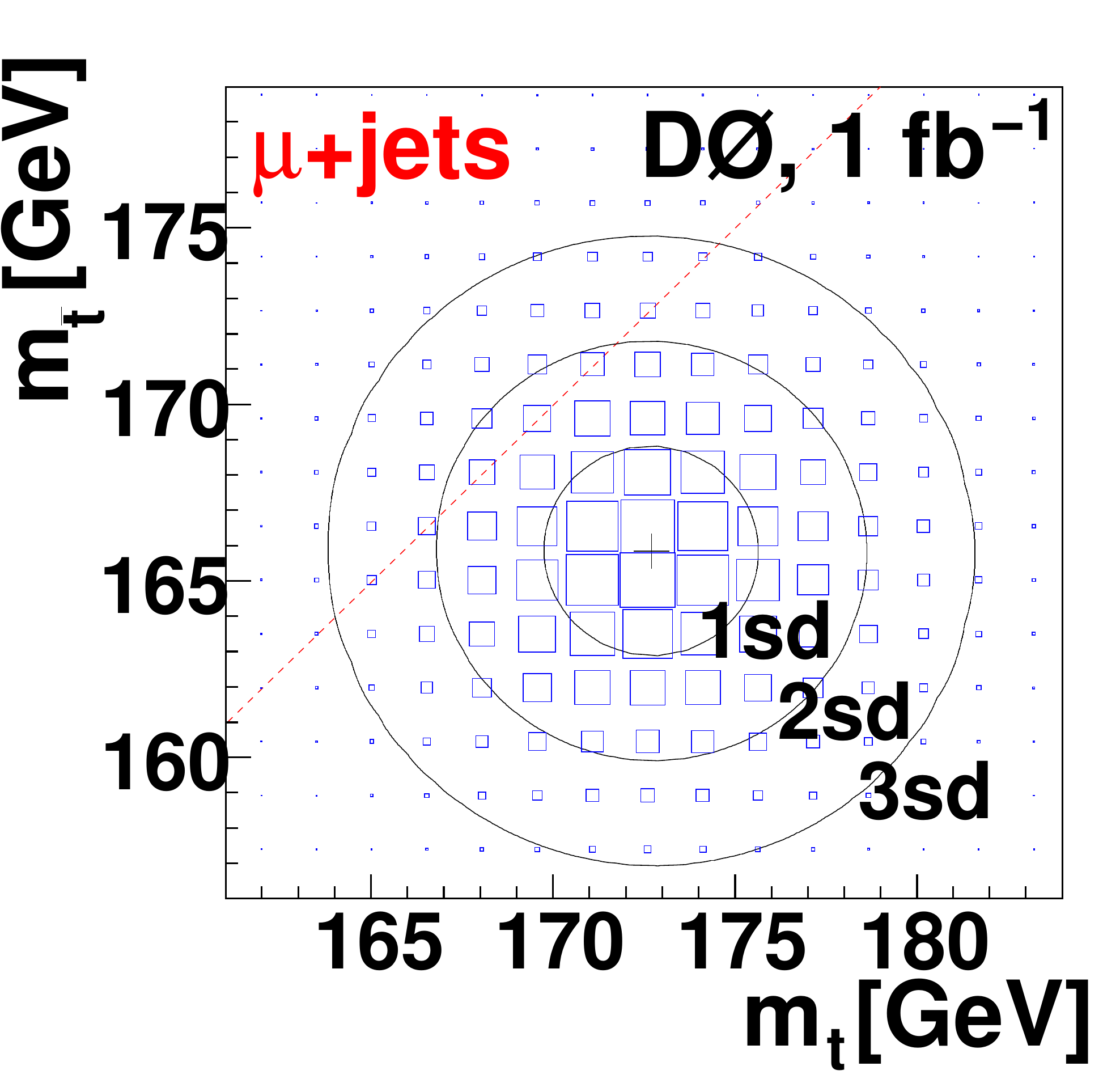}}%
\caption{Left: Calibration curve from simulated experiments
for the mass difference analysis
for the $e+\jets$ channel.  Middle and right: Fits to data.}
\label{mdiffres}
\end{figure}

Fig.~\ref{mdiffres} shows the analysis calibration
and fits to data.  Combining the electron and muon channels
yields $\Delta m = 3.8\pm3.4\ \textnormal{(stat.)}\gev$
and $\msum = 170.9\pm1.5\ \textnormal{(stat.)}\gev$.
The mass previously measured from this data sample was
$170.6\pm2.2\ \textnormal{(stat.+JES)}\gev$~\cite{mass1a}.

Systematic uncertainties are summarized
in Table~\ref{mdiffsyst}.  Most are common
with the mass measurement analysis; they tend, however, to cancel
in the difference.  Many systematics are dominated by the
statistics of the samples used to evaluate them.  Two
sources of systematics are new for this analysis.  First is the
uncertainty due to mismeasurements of the sign of the electron.
This is evaluated by increasing the mismeasurement rate in Monte Carlo
to match that seen in data.  Second is the asymmetry in the
calorimeter response to jets from $b$ and $\bar b$ quarks.  
This systematic is limited by statistics.

The total systematic uncertainty for the $\Delta m$ measurement is
$1.2\gev$, giving a final result from $1\ifb$ of data
of $\Delta m = 3.8\pm 3.7\gev$.


\begin{theacknowledgments}
This work is supported in part by the U.S. Department of Energy
under contract
DE-AC02-98CH10886 with Brookhaven National Laboratory.
\end{theacknowledgments}



\bibliographystyle{aipprocl} 

\bibliography{topmass}

\end{document}
